\documentclass[final, 3p, times, numbers, sort&compress]{elsarticle}
\usepackage{graphicx}
\usepackage{amsmath}
\usepackage{bm}
\usepackage[italicdiff]{physics}
\usepackage{wrapfig}
\usepackage{verbatim}
\usepackage{relsize}

\usepackage{color}
\usepackage{xcolor}
\usepackage{soul}
\usepackage{subcaption}
\usepackage[utf8]{inputenc}

\begin{document}

\journal{Computer Physics Communications}

\title{A portable and flexible implementation of the Wang--Landau algorithm in order to determine the Density of States}

\author[UNAB]{Felipe Moreno}
\ead{f.moreno.munoz@gmail.com}
\author[CCHEN,UNAB]{Sergio Davis}
\ead{sergio.davis@cchen.cl}
\author[UNAB]{Joaqu\'in Peralta}
\ead{joaquin.peralta@unab.cl}
\address[UNAB]{Departamento de F\'isica, Facultad de Ciencias Exactas, Universidad Andr\'es Bello, Sazi\'e 2212, Santiago, Chile.}
\address[CCHEN]{Comisi\'on Chilena de Energ\'ia Nuclear, Casilla 188-D, Santiago 7600713, Chile.}

\date{\today}

\begin{abstract}
In this work we develop an implementation of the Wang--Landau algorithm [Phys. Rev. Lett. \textbf{86}, 2050-2053 (2001)]. This algorithm allows us to find the density of states (DOS), a function that, for a given system, describes the proportion of states that have a certain energy. The implementation uses the Python and C++ languages for the algorithm itself, and it can take advantage of any library, such as the powerful LAMMPS library, for the computation of energy. Therefore, the resulting implementation is simple and flexible without sacrificing efficiency. This implementation also considers recent developments in the parallelization of the code for faster computation. We establish the soundness and effectiveness of our implementation by studying well-known systems such as the Ising model, the Lennard--Jones and EAM solids. We have found that our implementation can find the DOS
with very good precision in a reasonable amount of time. Therefore, we are equipped with a very powerful and flexible implementation that can be easily used in order to study more
realistic models of matter.
\end{abstract}
\maketitle
\noindent
{\bf PROGRAM SUMMARY}
 \\\\
\begin{small}
\noindent
{\em Program Title:} Republica Wang--Landau (RWL)\\
{\em CPC Library link to program files:} (to be added by Technical Editor) \\
{\em Code Ocean capsule:} (to be added by Technical Editor)\\
{\em Licensing provisions:} GPLv3  \\
{\em Programming language:} Python, C++ \\
  {\em Nature of problem:} An implementation of the WL algorithm that is flexible enough to be used for a large variety of systems.\\
  {\em Solution method:} This implementation separates the actual Wang--Landau code of the abstract implementation of the system. Therefore, any system can be attached as a walker---a Python class that represents the system.\\
  {\em Additional comments including restrictions and unusual features;} As examples, basic systems such as the Ising model are included plus a wrapper class for a LAMMPS walker to be used for any system supported by LAMMPS.\\

\end{small}
\section{\label{sec:Introduction}Introduction}

The density of states (DOS), denoted by $\Omega(E)$ and defined as $\Omega(E)=\int d\vb rd\vb p\,\delta(E-H(\vb r, \vb p))$, where $H(\vb r, \vb p)$ is the Hamiltonian of the system,
is a fundamental function of a system such that $\Omega(E)dE$ is the number of states that have an energy between $E$ and $E+dE$. This function can be used to compute all of the thermodynamic properties of a system. Examples of such properties are the caloric curve and the heat capacity, both being highly relevant to the study of a wide variety of systems. Therefore, computing the DOS is of fundamental importance, and consequently several methods for the calculation of this function for a given
system have been developed~\cite{Shimizu2004,Habeck2007,Montecinos2021}, some of them based on generalized ensembles and reweighting techniques.

Among the generalized-ensemble methods, Wang-Landau (WL) is a well known Monte Carlo technique for computing the DOS of a system~\cite{Wang2001}. This method was originally developed for discrete systems and having single processors in mind. However, since then the method has been expanded to off-lattice systems and parallel processing\cite{Vogel2013}. This has been done in order to take full advantage of the powerful multicore processing units that are the standard today. Although the WL algorithm is widely used and has a large number of variants and improvements~\cite{Liang2005, Atchade2010, Bornn2013}, there are still practical difficulties in its implementation, ranging from the fine-tuning needed for an optimal operation to the
lack of a universal implementation able to be ``plugged into'' traditional simulation codes.

In this work, we present a state-of-the-art implementation of WL using the flexibility and portability of the Python programming language~\cite{PYTHON}, coupled with the efficiency
of the C++ language. This implementation takes advantage of two well-known libraries: The powerful LAMMPS~\cite{Plimpton1995Mar} library, which will be used to compute the energy of
a complex configuration of atoms, and the easy-to-use Python multiprocessing library for parallel computing~\cite{Dalcin2011}. Furthermore, this implementation supports both discrete
and continuous systems, while either supporting or being ready to support established techniques to improve the algorithm such as the optimization of flatness criteria~\cite{Belardinelli2007}, parallelization scheme by replica-exchange~\cite{Vogel2013}, and more efficient Monte Carlo trial moves~\cite{Wust2009} among others. Some improvements
to the algorithm can be relatively easy to incorporate in the code, such as the WL-$1/t$ variant of Belardinelli et al.~\cite{Belardinelli2007}, which is already included in our code,
and or the Stochastic Approximation in Monte Carlo (SAMC) of Liang et al.~\cite{Liang2007}. Here we are presenting a well-modularized system that allows the scientific community to use
and incorporate different aspects of the WL algorithm efficiently.

The applicability of the WL algorithm is quite broad, covering different research areas such as multidimensional integration~\cite{Troster2005}, protein folding~\cite{Rathore2003a}, thermodynamics of polymers~\cite{Taylor2009}, free energy profiles~\cite{Calvo2002}, and the study of spin-glass models~\cite{Brown2005, Torbrugge2007, Alder2004}, among others. Here
we focus the technique mainly on the calculation of the caloric-curve for different cases, to expand its use towards the condensed matter community.
We tested this implementation using a variety of well-established benchmark systems such as the Ising model, the Potts model, classical and quantum harmonic oscillators, the
Lennard-Jones model and the embedded atom model (EAM). The goal of these tests was to prove the soundness of our implementation through the comparison with previously known results.

This paper is organized as follows. Following this introduction, in Section~\ref{sec:Algorithm} we describe the inner workings of the Wang--Landau algorithm. In Section~\ref{sec:Implementation} we offer a brief description of this implementation. In Section~\ref{sec:Simulations} we provide details of the benchmark simulations performed. A discussion of our findings is given in Section~\ref{sec:Conclusions}.

\section{\label{sec:Algorithm}A brief review of the Wang-Landau algorithm}
The WL algorithm is used to find the DOS of a many-particle system with $N$ degrees of freedom. This system can be described by a configurational state $X = (X_1,X_2,\ldots,X_N)$ and a function of $X$ that describes the system energy $E=\mathcal{H}(X)$. This function is called the Hamiltonian.

In order to produce the DOS of a system, the Wang--Landau algorithm takes advantage of the following property.
Suppose we sample a large number of possible states according to a distribution that depends only on the Hamiltonian of each state $P(X)=\rho(\mathcal{H}(X))$. If this distribution is chosen to be
\begin{equation}
\label{X-distribution}
    \rho(\mathcal{H}(X))\propto\frac{1}{\Omega(\mathcal{H}(X))},
\end{equation}
then the energies will be distributed according to
\begin{equation}
P(E)=\rho(E)\Omega(E)=\text{constant}.
\end{equation}
Consequently, we would build up a flat histogram. Thereby, we can use this property to test how good a candidate DOS is, and even better, we can determine how to make a small change to it so that it is closer to the true DOS. We keep on making those small changes to the candidate DOS until the histogram is flat enough. The family of algorithms that takes advantage of this property are known as flat histogram methods~\cite{landau2014}.

The specific steps to reproduce the above procedure for discrete systems is detailed now. At the start of our simulation we define a candidate DOS of the system, $\Omega(E)=1$ for all $E$. We define a \emph{walker} as an entity that wanders over the configuration space standing on an unique configuration at every simulation step. The energy of a walker is the energy of the configuration where this walker is currently standing on. We start out with a walker on a random state $X$, and then we produce a trial move from $X$ to $X'$. The transition acceptance that is in concordance with Eq. \ref{X-distribution} and the detailed balance condition is

\begin{equation}
    W(X\to X')=\min\left(1,\frac{\Omega(\mathcal{H}(X))}{\Omega(\mathcal{H}(X'))}\right).
\end{equation}
If the trial move is accepted, the walker moves to the new state $X'$; otherwise, it remains on its previous state $X$. Whether the change was accepted or not, we update the value of the DOS at the current energy $E$ according to $\Omega(E)\leftarrow \Omega(E)\cdot f$, where $f>1$. Based on the numeric nature of the algorithm, the values of these constants and their multiplication $\Omega(E)\leftarrow \Omega(E)\cdot f$ must be expressed in terms of their logarithms. Additionally, we update the histogram that counts the number of times certain energy has been visited. This process continues until the histogram is sufficiently flat. The flatness criterion is that the minimum value of the histogram must be at least 80\% of its mean value. Once this criterion has been met, the process is interrupted, the value of $f$ is decreased according to $f\leftarrow \sqrt{f}$, the histogram is reset, and the same process is started over again. Finally, the algorithm is stopped when $f$ reaches a value close enough to unity, such as $f\approx1+10^{-8}$, and the final $\Omega(E)$ is considered to be the true DOS.

For continuous systems, the energies are discretized into small bins of width $\Delta E$. Each bin has its own $\Omega$ and histogram, and they are updated when the energy of the walker lies between $E-\Delta E/2$ and $E+\Delta E/2$. The transition probability is computed after previously performing an interpolation on the discrete $\Omega$, thus obtaining $\Omega(E)$ from the interpolated function. This last step provides far better results than merely choosing the discrete $\Omega$ of the bin that contains $E$. This is because the walker will prefer a transition to a less probable energy, even if this new energy and the current one belong to the same bin. Furthermore, instead of the discrete case where the way in which we produce a change is obvious, we must select an appropriate step size, or alternatively, we can use a variable step size through the definition of a target acceptance. A target acceptance of 30\% has been used in the continuous systems.

\begin{figure}[t]
\centering
\includegraphics[scale=1]{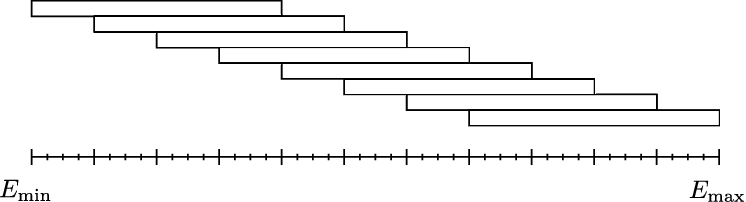}
\caption{The entire energy range is split up into overlapping windows.}
\label{fig:parallel}
\end{figure}

The algorithm is parallelized in the following way. Firstly, the entire energy range is split up into overlapping windows of energy (Fig.~\ref{fig:parallel}). An overlap of 75\% has been reported to provide good results~\cite{Vogel2013}. Secondly, every window is given its own walker and then a so-called ``sweep'' is executed---the performing of a certain number of simulation steps inside of every window. These steps run in parallel and they are completely independent. Thirdly, a replica exchange procedure~\cite{Vogel2013} is tried. This procedure consists in the exchange of walkers between two windows. For this exchange to occur, the windows must be adjacent and their walkers must have an energy that lies over the overlapping interval of their respective windows. If the previous conditions are met for windows $i$ and $j$, the exchange is performed according to the probability
\begin{equation}
P(w_i\leftrightarrow w_j)=\min\left(1,\frac{\Omega_i(E_i)\Omega_j(E_j)}{\Omega_i(E_j)\Omega_j(E_i)}\right),
\end{equation}
where $w$, $E$ and $\Omega$ are the walker, the walker's current energy and the current DOS of their respective windows. This procedure allow walkers to travel through the entire energy landscape thus avoiding the presence of energy traps. Fourthly, we repeat the last two steps until every window has ended its own simulation. The windows that have already ended their simulations are required to continue running them without updating their DOS. This is required because we need that those windows keep on performing replica exchanges with the active ones. Finally, we reconstruct the full DOS from the pieces of DOS obtained at every window, as follows.

In order to reconstruct $\Omega$, we must consider that we want to produce the smoothest curve possible, in particular, we want the microcanonical caloric curve $\beta(E)=\partial(\log\Omega(E))/\partial E$ to be smooth. Having this in mind, we select the pieces of $\Omega$ that belong to the first ($\Omega_1$) and second ($\Omega_2$) windows, and then we join them through the following procedure. Firstly, we compute $\beta_1(E)=\partial(\log \Omega_1(E))/\partial E$ and $\beta_2(E)=\partial(\log\Omega_2(E))/\partial E$ and we select the energy $E^*$ where the best match between these two functions is. In our second step, we define $\bar\Omega_2=[\Omega_1(E^*)/\Omega_2(E^*)]\Omega_2$ so that $\Omega_1$ and $\bar\Omega_2$ match their values at $E^*$. Thirdly, we join $\Omega_1$ and $\bar\Omega_2$ at $E^*$, cutting out the leftover ``tails'' of each function. Finally, the above steps are repeated between the previous pasted $\Omega$ and the next piece of DOS until there are no more pieces left. By this previous procedure, we can now join the DOS from the energy windows into a DOS that spans the whole energy range.

\section{\label{sec:Implementation}Implementation}

\begin{figure}[ht]
    \centering
    \includegraphics[scale=0.7]{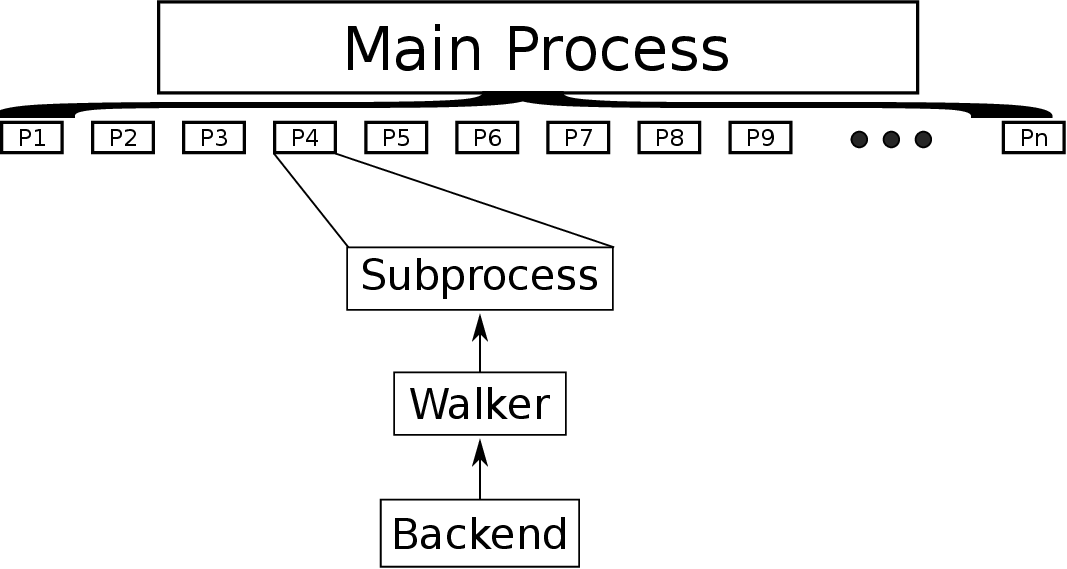}
    \caption{Diagram of the main process splitted into different subprocesses with a backend. }
    \label{fig:mainprocess}
\end{figure}

Before discussing the implementation details, we need to address some early issues. Firstly, we need to generalize the definition of the DOS to include discrete systems. For discrete systems $\Omega(E)$ will be simply the number of states with exact energy $E$. Secondly, to avoid treating with very large numbers, the logarithm of the DOS ($\log \Omega(E)$) will be computed instead of the plain DOS~\cite{Landau2004, Barash2017}. All equations will be transformed accordingly. Thirdly, after each sweep $\log\Omega(E)$ will be normalized such as that $\log\Omega(E_{\text{min}})=0$ or, equivalently, $\Omega(E)$ will be computed in units of $\Omega(E_{\text{min}})$.

This implementation is based on several modules. Firstly, we have a module that loads the simulation data such as the energy range, step size, target acceptance and so on from an input file. Secondly, there is the module that runs the actual simulations. This module spawns the subprocesses parallelization that will run the WL simulation on every window, performs the replica exchanges, merges the DOS pieces and writes the result into a file (Fig.~\ref{fig:mainprocess}). Finally, there are the walker modules, one for each system. These walkers keep track of the state of the system, perform changes, and compute the energy. Furthermore, those walkers are totally independent of the algorithm itself, so they can be used for any other algorithm and they can be coded in another language and wrapped into Python code. For example, a LAMMPS walker has been developed which contains a wrapper for the C++ LAMMPS library. Any other library can be used through a wrapper class. Because of this, this implementation is not affected by the lesser computational efficiency of an interpreted language such as Python, since the time consuming functions can be implemented in a faster language like C++ or Julia.

It is crucial to notice that in this current version, the parallelization is not based on the MPI library but in the Python \texttt{multiprocessing} module, which is designed for multi-core systems. The user has to look out for possible conflicts between this module and the parallelization of the walker in case it is an external process.

\subsection{Walkers}
Every Walker is a Python class that contains at least the following methods:
\begin{itemize}
    \item \texttt{energy()}: This function returns the current energy of the system. The function does not necessarily computes the energy---in some cases, it is useful to have it pre-computed beforehand.
    \item \texttt{change()}: This function makes a change in the system. The new energy can be computed immediately or when the previous function is invoked. No return value.
    \item  \texttt{undo()}: This function undo the last change performed by the above function. No return value.
    \item \texttt{state()/state(state):} These are a pair of setter and getter functions for the abstract representation of the state of the system. The first one set the state of the system and the second one returns the current state. This abstract representation can be any serializable Python object. These methods are needed in order to perform replica-exchanges.
\end{itemize}
An additional method can be used to set up an specific initial configuration for each simulation window:
\begin{itemize}
    \item \texttt{setup(index):} This method (if exists) will be invoked after the construction of the walker. An unique index representing the window that contains the walker will be passed as an argument. This way the user can perform specific tasks for each walker.
\end{itemize}
The implementation includes as examples some walkers for several common systems such as the Ising model, the Potts model and the harmonic oscillators (quantum and classical). They are thin wrappers over efficient C++ code. A LAMMPS walker is also included. This walker allows the simulation of any system supported by LAMMPS. Those walkers are not linked to a WL simulation in any way, so they can be easily imported and used for any other algorithm that complement an investigation such as the Metropolis--Hastings or Simulated Annealing algorithms.

There are two ways to run a simulation using this module. Firstly, the module can be imported into a Python script file where the user can create the main WL object and invoke the simulation with the \texttt{run()} method. Secondly, a script that loads the simulation from an input file is included.

\subsection{Running}
The algorithm starts with an user defined initial configuration. Next, the program will explore the energy landscape in order to find a set of configurations whose energies belong to every energy window. This is necessary because we need an initial configuration to start every subprocess. The algorithm used is very basic, it just explores the energy landscape randomly until it finds a suitable configuration. The user can provide her own previously defined configurations using the walkers \texttt{setup()} method. Otherwise, for best results, it is preferable to previously search for the lowest energy configuration using a more powerful algorithm, and to start the simulation using this configuration. This is because the lower energies are usually the most difficult to find. For LAMMPS systems, it is best to use the LAMMPS own minimization algorithms. Notice that if any energy window has an energy range that is impossible to arrive at from the initial configuration, the algorithm can not run.

\section{\label{sec:Simulations}Simulations and computational procedures}
A set of different simulations were performed in this work in order to determine the accuracy and practical usefulness of our implementation. For every system, instead of computing the full DOS, we only computed the configurational density of states (CDOS). This means that instead of using the full Hamiltonian $\mathcal{H}=K+\phi$, we only used the configurational part $\phi$. This is done because the contribution of the kinetic part $K$ is always the same and it is easy to compute analytically.

The caloric curve for both microcanonical and canonical ensembles has been computed using the CDOS ($\Omega(\phi)$) obtained in the simulation. If $d$ is the number of degrees of freedom in the system, the microcanonical curve has been computed according to

\begin{equation}
\ev{\beta}_E = \pdv{E}\log\eta(E),
\end{equation}
where
\begin{equation}
\eta(E) \equiv \int\Omega(\phi)\left[E-\phi\right]^{\frac{d}{2}-1}d\phi
\end{equation}
is a function that is proportional to the full DOS~\cite{Davis2020d}, and the canonical curve has been computed according to the well known equation
\begin{equation}
\ev{E}_\beta=\frac{d}{2\beta}-\pdv{\beta}\log Z(\beta),
\end{equation}
where
\begin{equation}
Z(\beta)\equiv\int \Omega(\phi)e^{-\beta \phi}d\phi
\end{equation}
is the configurational part of the partition function, and $d/2\beta$ is the kinetic energy term.

The results are presented below. The discrete models and the classical harmonic oscillator are presented as a proof of the soundness of our implementation. The Lennard--Jones and EAM solids are the most important results because they show how our code works seamlessly with the LAMMPS library which enables us to study a
great quantity of systems with high precision.

The number of steps per second reported for each system corresponds to standard simulations performed using 88 walkers running on an Intel(R) Xeon(R) CPU E5-2699 v4 @ 2.20GHz with 88 virtual cores on 44 physical cores.
\subsection{Discrete Models}

\begin{figure}
\begin{center}
\includegraphics[width=0.80\textwidth]{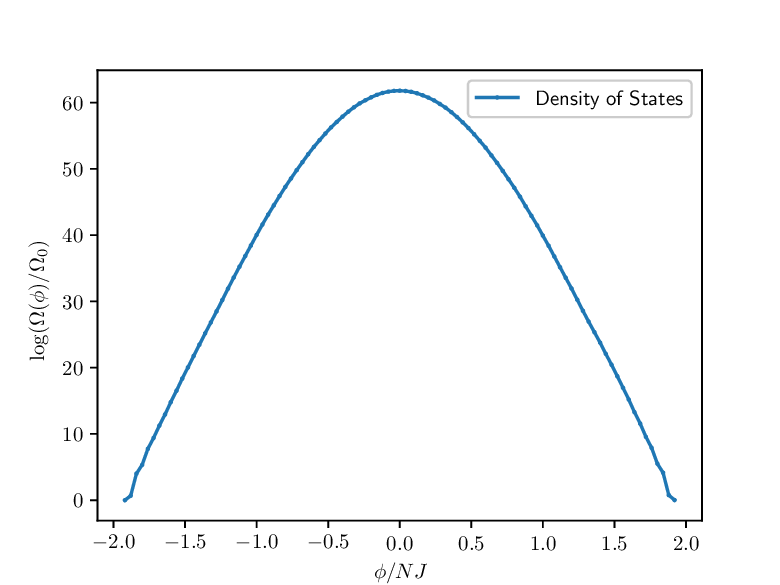}
\includegraphics[width=0.80\textwidth]{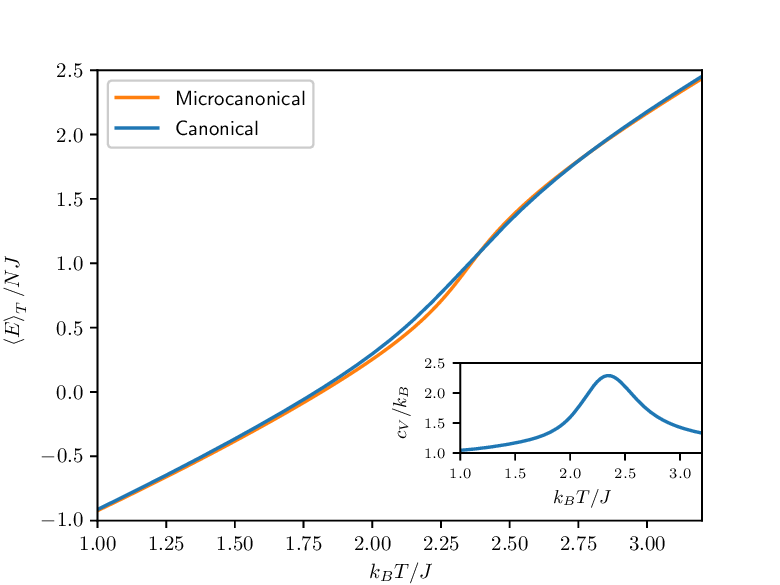}
\end{center}
\caption{CDOS (top) and caloric curve (bottom) for the 2-dimensional Ising model on a $10\times 10$ lattice. The inset shows the specific heat. Note that the caloric curve unusually contains the contribution of the kinetic energy of the system, only for purposes of comparison with atomistic models.}
\label{Ising-2D}
\end{figure}

The Ising model is the first discrete model studied, it is a well known model of interacting spins. The spins can take two values---up and down. This is one of the most simple models that exhibit a phase transition. The interaction between the spins is described by
\begin{equation}
  \phi = -J\sum_{\langle i,j\rangle} \bm{\sigma_i}\cdot\bm{\sigma_j},
\end{equation}
where the sum is over adjacent pairs. Usually, there is an additional term with the contribution of an external magnetic field. We use an 2D Ising Model of $N=100$ spins in a $10\times 10$ lattice without the magnetic field. The continuous transition can be observed in the caloric curve computed in the simulation shown in Fig.~\ref{Ising-2D}. The transition temperature is $k_BT_c/J=2.332$ which is close to the well-known exact transition temperature $k_BT_c/J = 2.269$ of this model~\cite{Baxter2016}. The number of steps per second for this system is $145733 \pm 44577$.

\begin{figure}
\begin{center}
\includegraphics[width=0.8\textwidth]{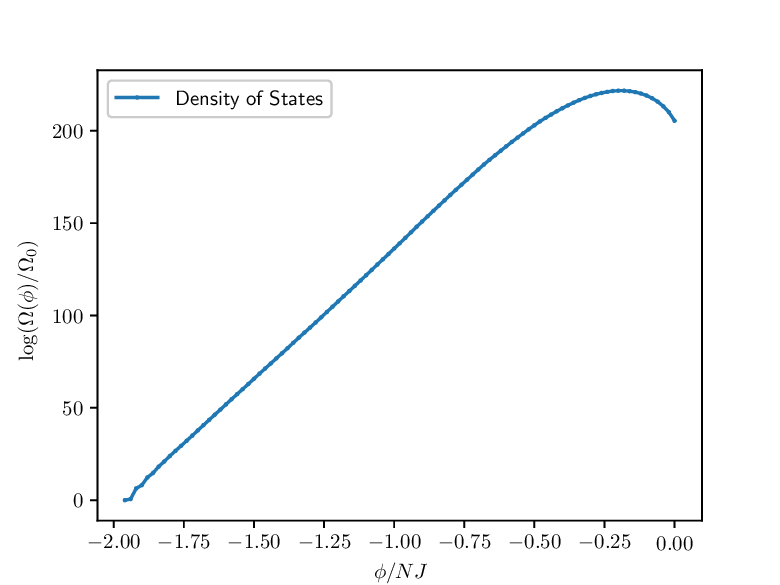}
\includegraphics[width=0.8\textwidth]{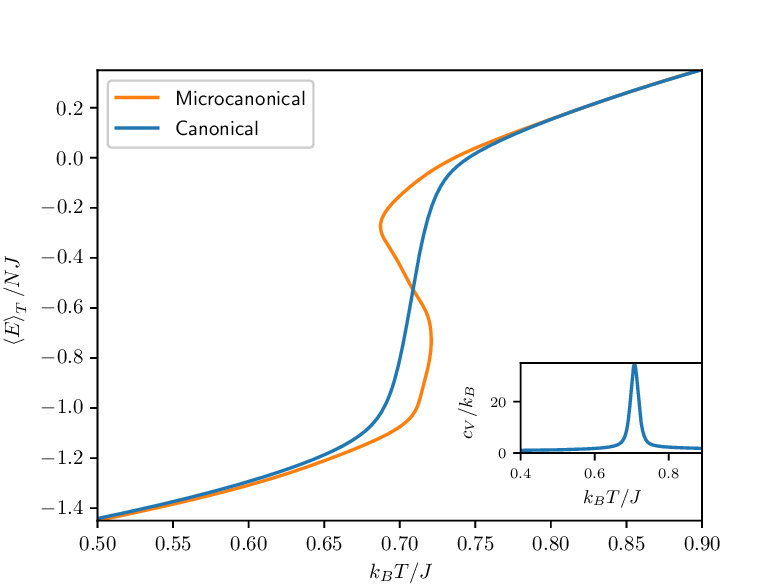}
\end{center}
\caption{CDOS (top) and caloric curve (bottom) for the Potts model with $q=10$ on a $10\times 10$ lattice. The inset shows the specific heat. Note that the caloric curve unusually contains the contribution of the kinetic energy of the system, only for purposes of comparison with atomistic models.}
\label{Potts}
\end{figure}

A more generalized discrete study case is the Potts model~\cite{Wu1982Jan}. This is a generalized Ising model of interacting classical spins $q$ on a regular lattice, which has been
widely used in multiple research areas~\cite{Turner2002, Davis2014, Farias2020, Torbrugge2007, Alder2004, landau2014}. In this model the spins can
take $q$ different ``orientations'', and its interaction energy is given by
\begin{equation}
\phi=-J\sum_{\langle i,j\rangle}\delta(\sigma_i,\sigma_j)
\end{equation}
where the sum is over adjacent pairs and $\delta$ is the usual Kronecker delta. In the particular case of $q=2$ we recover the 2D-Ising model up to a multiplicative factor. In what follows, we use a value of $q=10$ on a $10\times 10$ 2-dimensional lattice. This model exhibits a first-order phase transition which can be observed looking at the caloric curve obtained in the simulation (Fig.~ \ref{Potts}).

We obtained the same result than in our previous work where the phase transitions and metastable phases of this model have been characterized~\cite{Moreno2018}. The metastable region lies inside the S-shaped loop in the microcanonical curve. The number of steps per second for this system is $145142\pm 38669$.
\begin{figure}
\begin{center}
\includegraphics[width=0.8\textwidth]{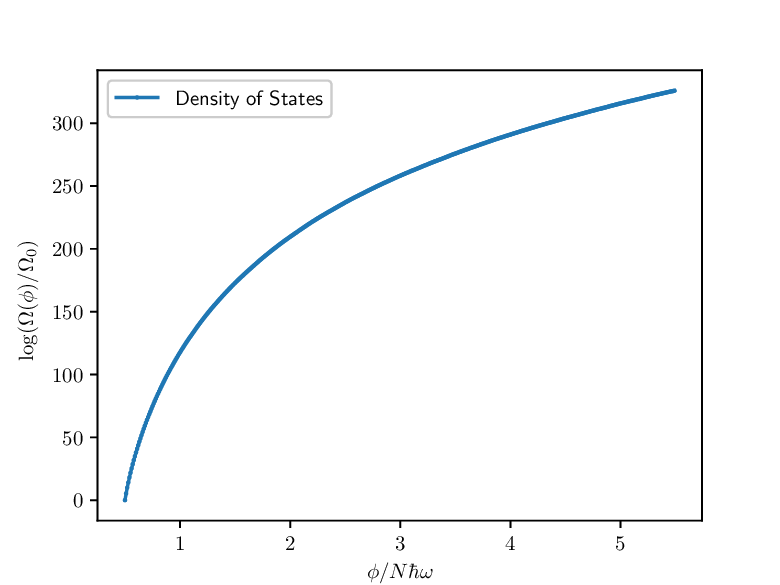}
\includegraphics[width=0.8\textwidth]{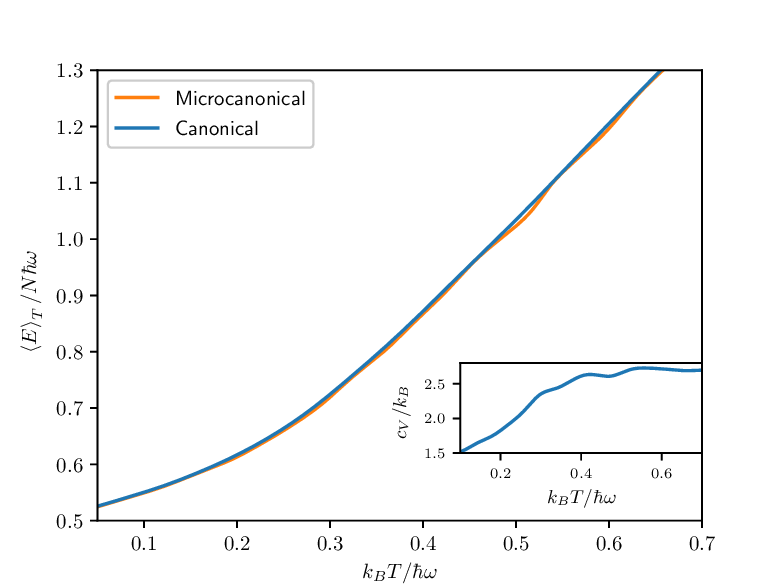}
\end{center}
\caption{CDOS (top) and caloric curve (bottom) for a system of 100 Quantum Harmonic Oscillators. The inset shows the specific heat.}
\label{qho}
\end{figure}

The third, and last, discrete model is a system of $N$ identical 1-dimensional quantum harmonic oscillators~\cite{Dekker1981, Chung2002} described by 
\begin{equation}
    \phi=\frac 12\hbar\omega\left(\sum_{i=1}^N n_i+\frac 12\right),
\end{equation}
where $n_i$ is a non-negative integer that is equal to the quantum number of the i-th oscillator. We consider $N=100$ oscillators. The plot (Fig.~\ref{qho}) shows how the caloric curve starts with a very low slope and then this slope gradually reaches the well-known~\cite{Greiner2012} classical value $E/Nk_BT=1$. The number of steps per second for this system is $154264\pm 28601$.

\subsection{Continuous Systems}

\begin{figure}
\begin{center}
\includegraphics[width=0.8\textwidth]{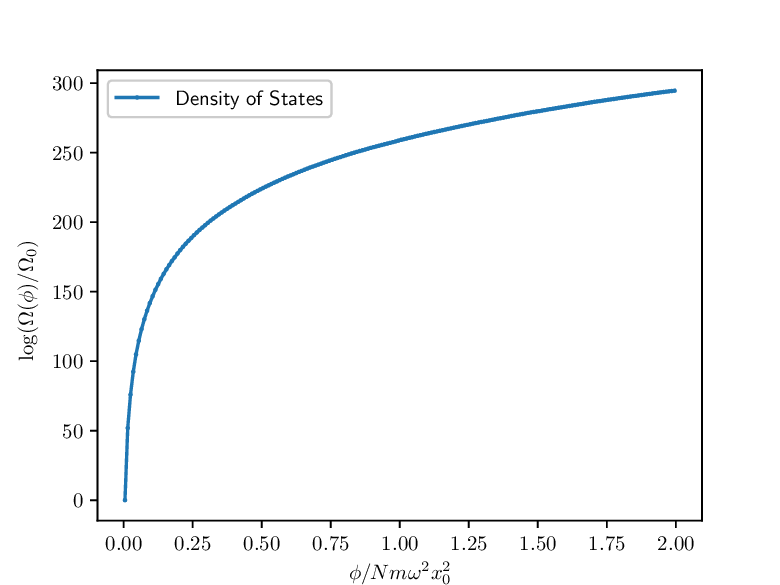}  
\includegraphics[width=0.8\textwidth]{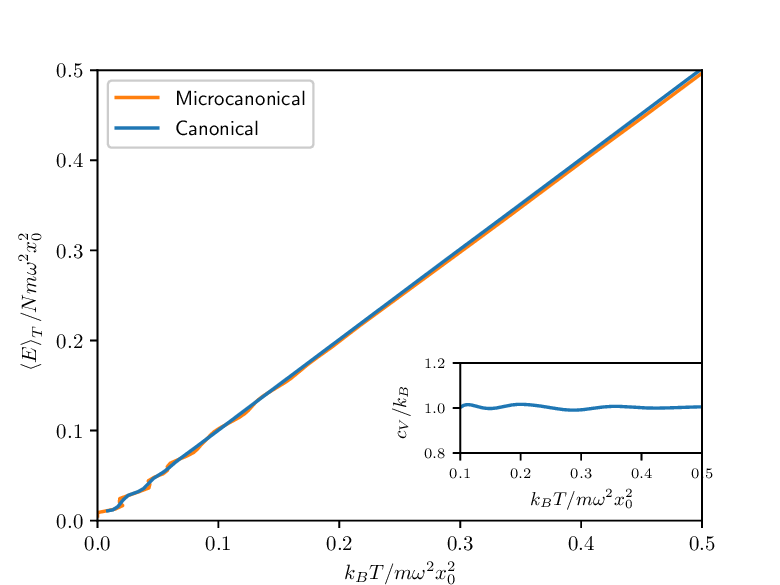}  
\caption{CDOS (top) and caloric curve (bottom) for a system of 100 Classical Harmonic Oscillators. The inset shows the specific heat.}
\label{ho}
\end{center}
\end{figure}

As a first example of a continuous system we consider a system of $N$ identical 1-dimensional classical harmonic oscillators described by
\begin{equation}
    \phi = \frac 12m\omega^2\sum_{i=1}^N x_i^2.
\end{equation}
where $x_i$ describes the position of the i-th oscillator. The energy is computed in units of $m\omega x_0^2$ where $x_0$ is an arbitrary length. At every step a random oscillator is displaced a random quantity drawn from an uniform distribution $[-x_0,x_0]$. As expected, in Fig.~\ref{ho} we obtain a straight line as the caloric curve~\cite{Greiner2012}, with the correct slope of 1. The number of steps per second for this system is $169008\pm 46026$.

\begin{figure}
\begin{center}
\includegraphics[width=0.8\textwidth]{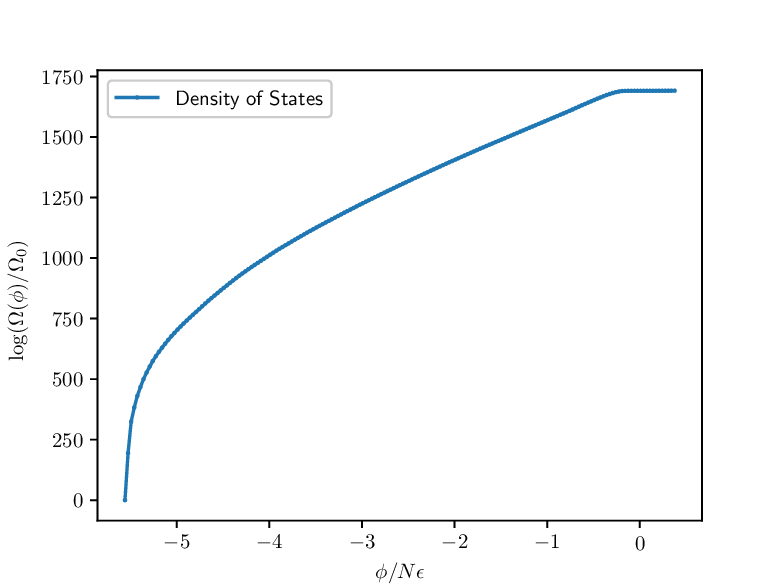} 
\includegraphics[width=0.8\textwidth]{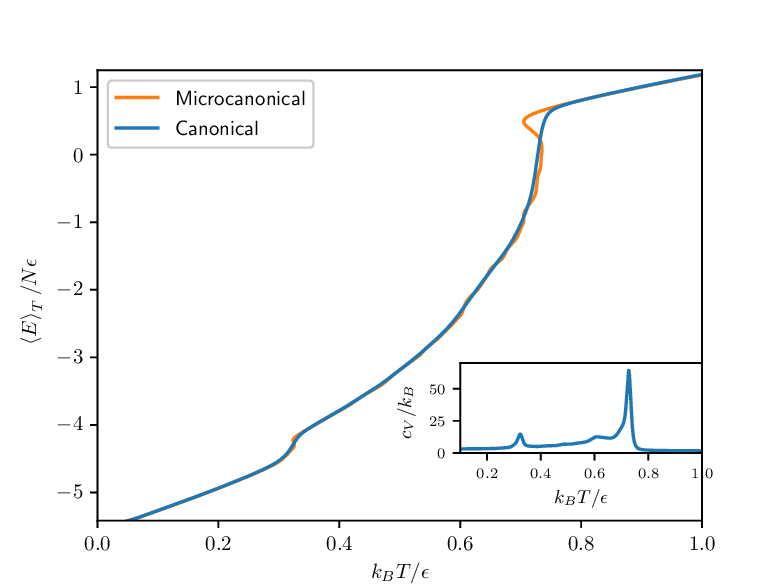} 
\end{center}
\caption{CDOS (top) and caloric curve (bottom) for a Lennard--Jones system with $N=120$ atoms. The inset shows the specific heat.}
\label{lj}
\end{figure}

The second continuous system studied is a Lennard--Jones (LJ) solid by using the energy provided by the classical LJ potential~\cite{lj1931}. The LJ potential is described by
\begin{equation}
    \phi=4\epsilon\sum_{i.j<i}\left[\left(\frac{\sigma}{r_{ij}}\right)^{12}-\left(\frac{\sigma}{r_{ij}}\right)^6\right].
\end{equation}
where the sum is over every pair of atoms. We utilize a cut-off value of $2.5\sigma$ which it means that we neglect the contribution of pairs where $r_{ij}>2.5\sigma$.  The structure is made of $N=120$ atoms with masses $m$ in a cubic cell of side $15\sigma$. This potential and the EAM potential are the most important model studied since they use the LAMMPS library. For this model, an initial minimization has been performed in order to find the low energy solid states, thus determining an appropriate energy range for our simulation. Looking at the caloric curve (Fig.~\ref{lj}) we observe a transition temperature of $k_BT/\epsilon\approx 0.7$ which agrees with the literature. This model produces very accurate results for noble gases, for example, for the Argon solid in which $\epsilon=1.03\times10^{-2}\text{ eV}$, this model reports a temperature of melting close to $T=83.59\text{ K}$. The number of steps per second for this system is $2301 \pm 277$.
\begin{figure}
\begin{center}
\includegraphics[width=0.8\textwidth]{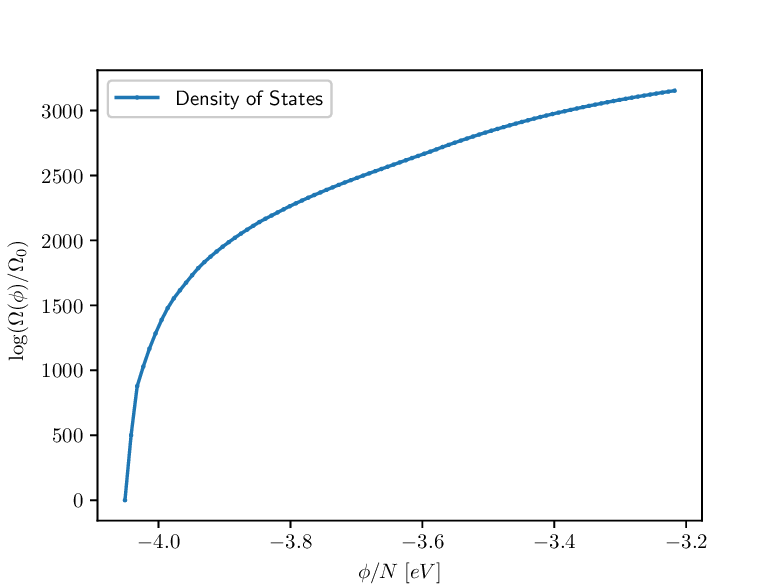}  
\includegraphics[width=0.8\textwidth]{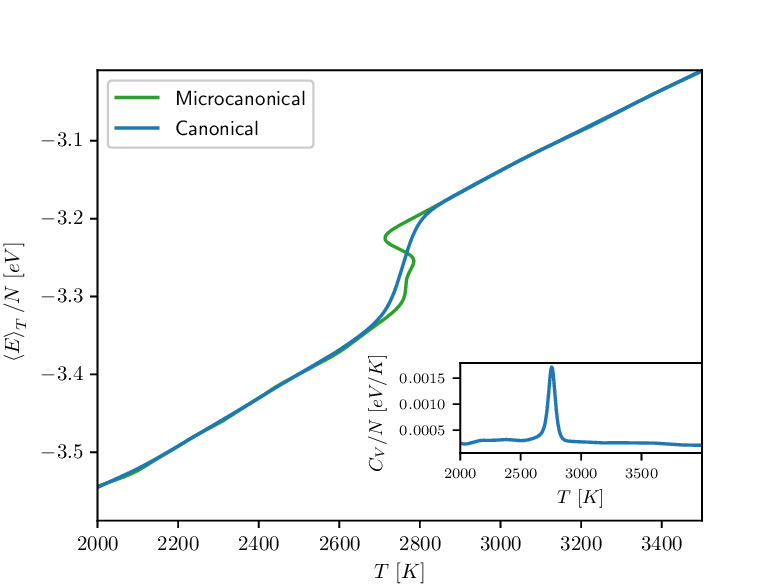}  
\end{center}
\caption{CDOS (top) and caloric curve (bottom) for an EAM system with $N=432$ atoms. The inset shows the specific heat.}
\label{eam}
\end{figure}

As a last study case, we use the embedded atom model (EAM) interatomic potential. The EAM is widely used in metallic systems, and incorporate a local density on the atomic interaction. A generalized form of this type of potential es given by

\begin{equation}
    E_i = F_\alpha\left(\sum_{i\neq j} \rho_\beta (r_{ij}) \right) + \frac{1}{2} \sum_{i\neq j} \phi_{\alpha\beta}(r_{ij})
\end{equation}
where $r_{ij}$ is the distance between atoms $i$ and $j$, $\phi_{\alpha\beta}$ is a pair-wise potential function, $\rho_\beta$ is the contribution to the electron charge density from atom $j$ of type $\beta$ at the location of atom $i$, and $F$ is an embedding function that represents the energy required to place atom $i$ of type $\alpha$ into the electron cloud. Even
when this kind of potentials have been widely used~\cite{daw1984,DAW1993251,ACKLAND2020544}, they are still highly time-consuming on computational procedures, and therefore they provide
a useful testing ground in order to define in an accurate way the scalability of our implementation of the WL-algorithm. The parameters used in the simulations are the ones that fit
the energies for Fe atoms~\cite{malerba2010}. The results are presented in Fig.~\ref{eam}. The WL algorithm has been tested on EAM models previously~\cite{Perera2016, ALEKSANDROV201079}, presenting an accurate estimation of the DOS. The number of steps per second for this system is $6.32\pm 0.12$.

\section{\label{sec:Conclusions}Concluding remarks}
Our work presents a modular implementation of the original Wang--Landau algorithm enhanced with state-of-the-art developments such as off-lattice systems and parallelization. It also integrates seamlessly with the LAMMPS library for fast computing of the energy of atomic configurations, which means that we can, in principle, study any system that can be described by a potential supported by LAMMPS, or, in principle, any other atomistic simulation package for which a wrapper can be written. The results obtained in our simulation of well known systems matched what we expected, including the specially important cases of the Lennard-Jones system and the EAM system because these are the models that run with the LAMMPS library. Consequently, our implementation of the WL algorithm can be used from now on to efficiently study the properties of more realistic models of matter.

\section*{Acknowledgments}
SD acknowledges partial support by ANID FONDECYT 1171127 and ANID PIA ACT172101 grants. FM thanks to Universidad Andrés Bello for providing the doctorate scholarship and acknowledges funding from ANID BECAS/DOCTORADO NACIONAL 21212267. Computational work was supported by the supercomputing infrastructures of the NLHPC (ECM-02), and FENIX (UNAB).

\bibliography{cdoswl}
\bibliographystyle{apsrev}

\end{document}